# Global texture in Lyra geometry

Farook Rahaman

Khodar Bazar, Baruipur – 743302
24-Parganas (South), West Bengal, India

E-mail : farook_rahaman@yahoo.com

Abstract:
  In this paper, we consider global texture with time dependent displacement vector based on Lyra geometry in normal gauge i.e. displacement vector $\varphi^*_i = (\beta(t), 0, 0, 0)$. We investigate gravitational field of global texture configuration by solving Einstein equations as well as that for the scalar field due to texture.

## Introduction:

Phase transitions in the early Universe can give rise to topological defects of various kinds. The appearance of global textures in the early Universe during a phase transition is predicted by Grand Unified Theories [1]. It is characterized by third order homotopy group ( in fact, $\pi_3(M) \neq I$, then texture will appear, M is the vacuum manifold ) [2]. Textures are stable, non localized solutions to the classical equations of a spatial manifold with compact dimension [ 2-3 ]. These structures collapse as soon as they come within the horizon i.e. when they become causally connected [2-4]. They suggested that this kind of topological defects might have been responsible for the formation of large scale structure .

In last few decades, there has been considerable interest in Alternative theories of gravitation. The most important among them are scalar-tensor theories proposed by Lyra [5] and by Brans-Dicke[5]. Lyra [5] proposed a modification Riemannian geometry by introducing a gauge function in to the structure less manifold that bears a close resemblance to Weyl's geometry. In general relativity, Einstein succeeded in geometrizing gravitation by identifying the metric tensor with the gravitational potentials. In the scalar-tensor theory of Brans-Dicke, on the other hand, scalar field remains alien to the geometry. Lyra's geometry is more in keeping with the spirit of Einstein's principle of geometrization, since both the scalar and tensor fields have more or less intrinsic geometrical significance.

In the consecutive investigations, Sen [6] and Sen and Dunn [6] proposed a new scalar tensor theory of gravitation and constructed an analog of the Einstein field equation based on Lyra's geometry which in normal gauge may be written as

$$R_{ik} - \tfrac{1}{2} g_{ik} R + \tfrac{3}{2} \phi^*_i \phi^*_k - \tfrac{3}{4} g_{ik} \phi^*_m \phi^{*m} = -8\pi G T_{ik} \qquad (1)$$

where, $\phi^*_i$ is the displacement vector and other symbols have their usual meaning as in Riemannian geometry.

Halford [7] has pointed out that the constant displacement field $\phi^*_i$ in Lyra's geometry play the role of cosmological constant $\Lambda$ in the normal general relativistic treatment. According to Halford the present theory predicts the same effects within observational limits, as far as the classical solar system tests are concerned, as well as tests based on the linearized form of field equations. Soleng [7] has pointed out that the constant displacement field in Lyra's geometry will either include a creation field and be equal to Hoyle's creation field cosmology or contain a special vacuum field which together with the gauge vector term may be considered as a cosmological term

Subsequent investigations were done by several authors in scalar tensor theory and cosmology within the frame work of Lyra geometry [8].

In recent, I have studied some topological defects namely monopole, cosmic string and domain wall in the frame work of Lyra geometry [9].

Our aim in this paper is look at global texture with time dependent displacement vector based on Lyra geometry.

## 2 . An over view on Lyra geometry:

Lyra geometry is a modification of Riemannian geometry which bears a close similarity to Weyl's geometry.

Lyra defined the displacement vector PF between two neighboring points $P(x^i)$ and $F(x^i + dx^i)$ by its components $A\,dx^i$ is a gauge function A together form a reference system $(A, x^i)$.

The tranformation to a new reference system $(x^i, x_1^i)$ are given by

$$A_1 = A_1(A, x^i), \quad x_1^i = x_1^i(x^i) \qquad \ldots\ldots(2)$$

with $\partial A_1 / \partial A \neq 0$ and the Jacobian $|\partial x_1^i / \partial x^i| \neq 0$

The connections is taken as

$$^*\Gamma_{bc}^{\,a} = A^{-1} \Gamma_{bc}^{\,a} - \tfrac{1}{2}(\delta_b^{\,a} \varphi_c + \delta_c^{\,a} \varphi_b - g_{cb} \varphi^a) \qquad \ldots(3)$$

where, the $\Gamma_{bc}^{\,a}$ are defined in terms of the metric tensor $g_{ab}$ as in Riemannian geometry and $\varphi_a$ is a displacement vector field. Lyra [5] and Sen[6] have shown that in any general system the vector field quantities $\varphi_a$ arise as a natural consequence of the introduction of the gauge function A into the structure less manifold. $^*\Gamma_{bc}^{\,a}$ are symmetric in their lower two indices.

The metric in Lyra's geometry is given by

$$ds^2 = A^2 g_{ij} x^i x^j \qquad \ldots\ldots(4)$$

and is invariant both co ordinate and gauge transformations.

The infinitesimal parallel transport of a vector $\xi^a$ is given by

$$d\xi^a = -\tilde{\Gamma}_{bc}{}^a \xi^b A\, dx^c \qquad \ldots\ldots(5)$$

where,

$$\tilde{\Gamma}_{bc}{}^a = {}^*\Gamma_{bc}{}^a - \tfrac{1}{2}\delta_b{}^a \varphi_c \qquad \ldots\ldots(6)$$

The $\tilde{\Gamma}_{bc}{}^a$ are not symmetric in b and c .
The length of a vector does not change under parallel transport.
The curvature tensor is defined by

$$^*R^a{}_{bcd} = A^{-2}[-(A\tilde{\Gamma}_{bc}{}^a)_{,d} + (A\tilde{\Gamma}_{bd}{}^a)_{,c} - A^2(\tilde{\Gamma}_{bc}{}^e \tilde{\Gamma}_{ed}{}^a - \tilde{\Gamma}_{bc}{}^e \tilde{\Gamma}_{ce}{}^a)] \qquad \ldots(7)$$

The curvature scalar, obtained by contraction of eq.(7) is

$$^*R = A^{-2} R + 3 A^{-1} \varphi^a{}_{;a} + (3/2) \varphi^a \varphi_a + 2 A^{-1}(\log A^2)_{,a} \varphi^a \qquad \ldots\ldots(8)$$

The invariant volume integral is given by

$$I = \int L \sqrt{(-g)}\, A^4 d^4x \qquad \ldots\ldots(9)$$

where $d^4x$ is the volume element and L is a scalar invariant .
If we use a normal gauge i.e. $A = 1$ and $L = {}^*R$ in eq.(9) , then eqs.(8) and (9) become, respectively

$$^*R = R + 3\varphi^a{}_{;a} + (3/2) \varphi^a \varphi_a \qquad \ldots\ldots(10)$$

$$I = \int {}^*R \sqrt{(-g)}\, d^4x \qquad \ldots\ldots(11)$$

The field equations are obtained from the variational principle

$$\delta(I + J) = 0 \qquad \ldots\ldots(12)$$

where, I is given as by eq.(11) and J is related to the Lagrangian density **L** of matter by

$$J = \int \boldsymbol{L} \sqrt{(-g)}\, d^4x \qquad \ldots\ldots(13)$$

The field equations are thus [ using $\chi = (8\pi G/c^4)$ ]

$$R_{ik} - \tfrac{1}{2} g_{ik} R + (3/2) \phi_i \phi_k - \tfrac{3}{4} g_{ik} \phi_m \phi^m = -\chi T_{ik} \qquad \ldots\ldots(14)$$

# 3. Basic Equations:

A texture configuration can be described by the action integral [3,4] :

$$A = \int d^4x \, (\sqrt{-g}) [\, \tfrac{1}{2} \partial_\mu \phi^a \partial^\mu \phi^a - \tfrac{1}{4} \lambda (\phi^a \phi^a - \eta^2)^2 - (1/16\pi G) \,] \qquad \ldots\ldots(15)$$

Here $\phi^a$, a four component scalar field has the form [3,4] :

$$\phi^a = [\, \cos\chi, \, \sin\chi \sin\theta \cos\varphi, \, \sin\chi \sin\theta \sin\varphi, \, \sin\chi \cos\theta \,] \qquad \ldots\ldots(16),$$

for a spherically symmetric configuration of the texture, where $\theta$ and $\varphi$ are the usual spherical angular co ordinates and $\chi = \chi(r,t)$ with $\chi \to 0$ as $r \to 0$ and $\chi \to \pi$ as $r \to \infty$.
This model has an $0(4)$ which is spontaneously broken to $0(3)$ due to phase transition.
In fact the vacuum manifold is characterized by the 3-sphere $\phi^a \phi_a = \eta^2$.
When, $\phi^a$ acquires a vacuum expectation value, the residual symmetry is $0(3)$.
Therefore the relevant homotopy group that tells us that there are non trivial texture configuration is $\pi_3(\,0(4)/0(3)\,) = Z$. In this paper we shall consider a texture configuration with winding number unity.
We consider a global texture configuration with the metric ansatz

$$ds^2 = A(r,t) \, dt^2 - B(r,t) \, dr^2 - r^2 H \, d\Omega_2^2 \qquad \ldots(17)$$

The energy momentum tensor for the texture configuration is

$$T_{ij} = \nabla_i \phi^a \nabla_j \phi_a - [\, \tfrac{1}{2} (\nabla_k \phi^a)(\nabla^k \phi_a) \,] \, g_{ij} \qquad \ldots\ldots(18)$$

Now the explicit expression for the coupled Einstein and scalar field equation based on Lyra geometry (with the texture configuration given in eq.(16)) are

$$(1/Br^2) - (1/Hr^2) - (B'/r B^2) - \tfrac{1}{2} (B^\bullet H^\bullet / ABH) - \tfrac{1}{4} (H^{\bullet 2}/H^2 A) - \tfrac{3}{4} (\beta^2/A)$$
$$= K [\, (\chi'^2/2B) + (\chi^{\bullet 2}/2A) + (\sin^2\chi / r^2 H) \,] \qquad \ldots(19)$$

$$(1/Br^2) - (1/Hr^2) + (A'/rAB) + \tfrac{1}{2} (A^\bullet H^\bullet / HA^2) + \tfrac{1}{4} (H^{\bullet 2}/H^2 A) - (H^{\bullet\bullet}/HA) + \tfrac{3}{4} (\beta^2/A)$$
$$= K [\, -(\chi'^2/2B) + (\chi^{\bullet 2}/2A) + (\sin^2\chi / r^2 H) \,] \qquad \ldots(20)$$

$$\tfrac{1}{2} (1/AB) [\, -B^{\bullet\bullet} + (A'/r) + A^{11} + (B^{\bullet 2}/2B) + (A^\bullet B^\bullet /2A) - (A'B'/2B) - (B^\bullet H^\bullet /2H) \,]$$
$$- (B'/2rB^2) - (A'/4BA^2) + \tfrac{1}{4}(H^{\bullet 2}/H^2 A) - (H^{\bullet\bullet}/2HA) + (A^\bullet H^\bullet /4HA^2) + \tfrac{3}{4}(\beta^2/A)$$
$$= K [\, (\chi'^2/2B) - (\chi^{\bullet 2}/2A) \,] \qquad \ldots\ldots(21)$$

$$\tfrac{1}{2} [\, -(H^\bullet /Hr) + (B^\bullet /rB) + \tfrac{1}{2}(A'H^\bullet/AH) \,] \;=\; K \chi' \chi^\bullet \qquad \ldots\ldots(22)$$

[ heree, $K = 8\pi G \eta^2$ and dot and prime denote the differentiation w.r.t. t and r respectively.]

The equation of motion for $\varphi^a$, $\nabla^j \nabla_j \phi^a = -[\{(\nabla_j \phi_b)(\nabla^j \phi^b)\}/\eta^2] \phi^a$
becomes for the above metric [3,4]

$$(\chi^{11}/B) - (\chi^{\cdot\cdot}/A) + (\chi^1/B)[(2/r) - (B^1/2B) + (A^1/2A)] + (\chi^{\cdot}/A)[(H^{\cdot}/H) - (B^{\cdot}/2B)$$

$$+ (A^1/2A)] = \sin 2\chi / r^2 H \qquad \ldots(23)$$

## 4. Solutions:

To solve the field equations we assume separable form of the metric coefficients namely

$$\chi = \chi_1(r) + \chi_2(t) \ ; \ A = A_1(r) A_2(t) \ ; \ B = B_1(r) B_2(t) \qquad \ldots\ldots(24)$$

Let further assume $\chi^{\cdot}_2(t) = a_0(H^{\cdot}/H)$ ; $\chi_1^1(r) = c(A_1^1/A_1)$ $\qquad \ldots\ldots(25)$

where, $a_0$ and c are arbitrary constants.

Then from the field eq.(22), we get,

$$B_2(t) = B_0 H^m \ ; \ A_1(r) = A_0 r^d \qquad \ldots(26)$$

where $A_0$ and $B_0$ are integration constants and m is the separation constant and

$$d = [(1-m)/(½ + 2 a_0 c K)] \qquad \ldots\ldots(27)$$

Using the above separable form (24) for the metric coefficients in the field equations (19) – (22), we have

$$B_1(r) = [1 r^{2-d} + g r^{-e}]^{-1} \ ; \ \chi_1(r) = c \ d \ \ln(r/r_0) \qquad \ldots\ldots(28)$$

$$A_2(t) = [H^{\cdot 2}/\{C_2 H^{-h} - a_2 H^{2-m}\}] \ ; \ \chi_2(t) = a_0[\ln(H/H_0)] \qquad \ldots\ldots(29)$$

$$\beta^2 = 2n H^{\cdot 2}[(l/m-1)/\{3 B_0 C_2 H^{m-h} - 3 B_0 a_2 H^2\}] - (m+a_1+1)(H^{\cdot 2}/3 H^2) \ \ldots(30)$$

where $r_0$, $H_0$, $C_2$ and g are integration constants and h, l, e, n, $a_2$ are constants.
Thus time part of the metric coefficients $A_2(t)$, $B_2(t)$ and the displacement vector $\beta^2$ can be expressed in terms of H(t). So for arbitrary H(t), one can find the exact solutions of the field equations.

# 5. Weak field Approximation:

In this section,, we are discussing a case by taking a particular value of H(t), $H(t) = e^{2t}$. Further, we assume $\chi$ is a function of r only, say $\chi = \chi(r)$ and displacement vector is a constant.
Since $\chi$ is a function of r only, then from eq.(22), we get,

$$- (H^{\cdot}/H\, r) + (B^{\cdot}/rB) + \tfrac{1}{2}(A^1 H^{\cdot}/AH) = 0 \qquad \ldots\ldots(31)$$

Hence we get,
$$B_2(t) = B_{00} H^q ; \quad A_1(r) = A_{00}\, r^{\,2(1-q)} \qquad \ldots(32)$$

[$A_{00}$ and $B_{00}$ are integration constants and q is a separation constant ]
Now if we ignore gravity, the equation for $\varphi^a$ results ( Barriola and Vachaspati )(BV) [4]

$$- \chi^{11} + (\chi^{\cdot\cdot}/A) - 2(\chi^1/r) = - (\sin 2\chi / r^2) \qquad \ldots(33)$$

BV assumed U = (r/t), V = t and $\chi$ is a function of U only. Then eq.(33) transforms to

$$\chi^{11} + 2(\chi^1/U) = [\sin 2\chi /\{U^2(1 - U^2)\}] \qquad \ldots(34)$$

The solution of the eq.(34) is [4]

$$\chi(r, t) = 2 \tan^{-1}(-r/t) \quad t < 0.$$
$$= 2 \tan^{-1}(r/t) \quad t > r > 0. \qquad \ldots\ldots(35)$$
$$= 2 \tan^{-1}(t/r) \quad r > t > 0.$$

If we choose the separation constant q = 1, then

$$B_2(t) = H ; \quad A_1(r) = 1 \qquad \ldots(36)$$

[ taking integration constants to be unity ]

Now assuming $A_2(t) = H = e^{2t}$, we get from the field equations

$$B_1(r) = [(1 + \tfrac{1}{2} K r^2 \chi^{1\,2})/(1 + a r^2 + K \sin^2 \chi)] \qquad \ldots\ldots(37)$$

[ where $a = (1 - \tfrac{3}{4}\beta^2)$ ]

So using the above solutions, the eq.(23) for $\chi$ results

$$r^2 \chi^{11}(1 + a r^2 + K \sin^2 \chi) + K r^3 (\chi^1)^3 (1 + 3a r^2 + \sin^2 \chi) - \tfrac{1}{2} K r^2 (\chi^1)^2 \sin 2\chi$$
$$+ r\chi^1 (2 + 3a r^2 + K\sin^2 \chi) = \sin 2\chi \qquad \ldots\ldots(38)$$

Thus for K = 0 ( i.e. in the absence of the texture ) the line element (17) becomes

$$ds^2 = e^{2t} [ dt^2 - ( 1 + r^2 )^{-1} dr^2 - r^2 d\Omega_2^2 ]  \qquad \ldots(39)$$

Let us take the following coordinates transformations

$$T = e^t [ 1 + r^2 ]^{1/2} \quad ; \quad R = e^t \cdot r \qquad \ldots(40)$$

Then the differential equation (38) becomes

$$\chi^{11}[1 + (a-1)U^2 + K(1-U^2)\sin^2\chi] + K U (1-U^2) (\chi^1)^3[1+ (3a-1)U^2 +(1-U^2)\sin^2\chi ]$$

$$- \tfrac{1}{2} K(1-U^2) (\chi^1)^2 \sin 2\chi + (\chi^1/U)[2 +3(a-1)U + K(1-3U^2)\sin^2\chi] = [\sin 2\chi /\{U^2(1-U^2)\}]$$

$$\ldots\ldots(41)$$

where, U = (R /T) and $\chi$ is a function of U only.

One can note that if we set K = 0, then eq.(41) will coincide with eq.(34) ( K= 0 , i.e. in the absence of texture we neglect the displacement vector ).
Thus we may conclude that the flat space solution for texture is self similar.
Further in the region $1<< U << (1/\sqrt{K})$ , the equation (41) can be written effectively as

$$( a - 1) U^2 \chi^{11} - K( 3a - 1) U^5 (\chi^1)^3 + ( a - 1) (\chi^1/U) = 0 \qquad \ldots\ldots(42)$$

Which has a first $\chi^1 = ( e^F / M )^{1/2} \qquad \ldots\ldots(43)$

Where F = (6 / U) and $M = 6^4 S [- (e^F / 4F^4) - (e^F / 12F^3) - (e^F /24F^2) - (e^F / 24F)$

$$+ ( 1 / 24 ) \{ \ln F + (F / 1.1! ) + (F^2 / 2.2! ) + (F^3 / 3.3! ) + \ldots\ldots\}] + D.$$

[ D is an integration constant and $S = 2K \{( 3a - 1 ) / (a - 1)\}$ ]

Hence the implicit expression of $\chi$ is

$$\chi = \int ( e^F / M )^{1/2} dU + E \qquad \ldots(44)$$

Though we can not get the exact analytical form of $\chi$ ,but one can see that the solution is self similar form .

# 6. Concluding Remarks:

In this paper we have studied gravitational field of a global texture in Lyra geometry.

We see that in the weak field approximation ( K << 1 ), the metric coefficients $B_1(r)$ takes the form ( from eq.(37) )

$$B_1(U) = [(1 - U^2) / \{1 + (a - 1) U^2\}] \quad \ldots\ldots(45)$$

Which for U >> 1 and fixed t gives the line element

$$ds^2 = -(4/3\beta^2) dr^2 - r^2 d\Omega_2^2 \quad \ldots(46)$$

This is just a conical metric with a deficit angle $2\pi [1 - (4/3\beta^2)]$.
Thus deficit angle for the texture in the weak field depends on the displacement vector.
For null radial geodesic, we set $ds^2 = d\Omega_2^2 = 0$ in the line element with metric coefficient (45) [4].
This gives,

$$(dr/dt) = [(1 - \tfrac{3}{4} \beta^2) / (1 - U^2)]^{\frac{1}{2}} \quad \ldots\ldots(47)$$

Hence we get,

$$|t| = |t_0| \exp[\check{u}\int^U dU [\{(1 - \tfrac{3}{4} \beta^2 U^2) / (1 - U^2)\}^{\frac{1}{2}} - U]^{-1} \quad \ldots(48)$$

where $t_0$ and Ŭ refer to some initial time and position respectively.
In the weak field case,

$$|t| = |t_0| [(\sqrt{3}/2\, \beta - U_0) / (\sqrt{3}/2\, \beta - U)] \quad \ldots\ldots(49)$$

[ In the region $U^2 >> 1$ ]

For future work it will be interesting to study different properties for texture in Lyra geometry.

## Acknowledgement:

I am thankful to Prof. S Chakraborty and Dr. S. Chatterji for helpful discussion. I wish to thank the UGC, Government of India for financial support.